\documentclass[aps,twocolumn,showpacs,pra]{revtex4}
\usepackage{epsfig, psfrag}
\usepackage{amssymb,amsmath}
\usepackage{color}

\newcommand{\bra}[1]{\left\langle#1\right|}
\newcommand{\hc}{\mathrm{h.c.}}
\newcommand{\idh}{\frac{i}{\hbar}}
\newcommand{\ket}[1]{\left|#1\right\rangle}
\newcommand{\littlemathskip}{~}
\newcommand{\tr}{\mathrm{tr}}
\newcommand{\Hcond}{H_{\mathrm{cond}}}

\newcommand{\ground}{\ket{\downarrow}}
\newcommand{\groundbra}{\bra{\downarrow}}
\newcommand{\excited}{\ket{\uparrow}}
\newcommand{\excitedall}{\ket{\uparrow_1 \ldots \uparrow_D}}
\newcommand{\excitedbra}{\bra{\uparrow}}
\newcommand{\nodetect}{\ket{0}}

\begin{document}

\title{Passage-time distributions from a spin-boson detector model}

\author{Gerhard C. Hegerfeldt} \author{Jens Timo Neumann}
\affiliation{Institut f\"ur Theoretische Physik, Universit\"at
  G\"ottingen, Friedrich-Hund-Platz 1, 37077 G\"ottingen, Germany}
\author{Lawrence S.\ Schulman}
\affiliation{Physics Department, Clarkson University, Potsdam, New
  York 13699-5820, USA}
\date{October 6, 2006}

\begin{abstract}
  The passage-time distribution for a spread-out quantum particle to
  traverse a specific region is calculated using a detailed quantum
  model for the detector involved. That model, developed and
  investigated in earlier works, is based on the detected particle's
  enhancement of the coupling between a collection of spins (in a
  metastable state) and their environment. We treat the continuum
  limit of the model, under the assumption of the Markov property, and
  calculate the particle state immediately after the first detection.
  An explicit example with 15 boson modes shows excellent agreement
  between the discrete model and the continuum limit. Analytical
  expressions for the passage-time distribution as well as numerical
  examples are presented. The precision of the measurement scheme is
  estimated and its optimization discussed. For slow particles, the
  precision goes like $E^{-3/4}$, which improves previous $E^{-1}$
  estimates, obtained with a quantum clock model.
\end{abstract}


\maketitle

\section{Introduction}

Time-of-flight measurements are a standard tool for many
experimentalists. Since the particles or atoms involved are usually
fast, their center-of-mass motion is typically treated classically,
yielding a simple description of the time-of-flight measurement. But
as the diffraction and interference experiments using temporal
(instead of spatial) slits of Szriftgiser et.\ al have shown
\cite{sgad1996}, such a description of the center-of-mass motion by
means of classical physics is not always sufficient: the advance of
cooling techniques has made it possible to create ultracold gases in a
trap and produce very slow atoms, e.g., by opening the trap. Whenever
such ultracold atoms are involved, the spatial extent and the
spreading of the wave function can show noticeable effects. Even the
seemingly simple question of the time spent by a particle in a given
region of space does not possess a simple and definite answer. Related
to this ``dwell-time'' problem are the problems of ``passage time,''
concentrating on those particles that actually cross the region of
interest and are not reflected, and ``tunneling time'' in the case of
a barrier that classically cannot be traversed inside the region of
interest. These problems have on one hand been treated axiomatically
aiming at ideal quantities relying only on the system of interest,
see, e.g., Refs.\ \cite{fts1960, sb1987, jw1988, mbs1992}. Other
approaches may be called ``operational'' in the sense that a sort of
``measurement device'' is introduced to which the system of interest
is coupled; see, e.g., Refs.\ \cite{ab1967, vr1967, bd1997, sw1958,
  ap1980, amm2003}. For a critical review on different approaches to
tunneling times see, e.g., Refs.\ \cite{hs1989, lm1994}.

A distinction that can be drawn between the various time-related
quantities recalled above concerns whether they are pre- or post-
decoherence. Since the present work models the measurement apparatus,
it can be thought of as post-decoherence; in fact it is one of our
objectives to include decoherence-inducing processes, thereby
eliminating some of the black magic of quantum measurement. On the
other hand, one can take, for example, a path integral approach to
tunneling time \cite{ziolkowski, sokolovski} in
which the paths are sorted by (variously defined) times spent in the
barrier region. For these paths, amplitudes are retained, so it is
very much a pre-decoherence calculation. As a result interpretive
issues arise (discussed for example by \cite{yamada}) which do not
enter in the present work.

\begin{figure}[ht]
\begin{center}
\epsfig{file=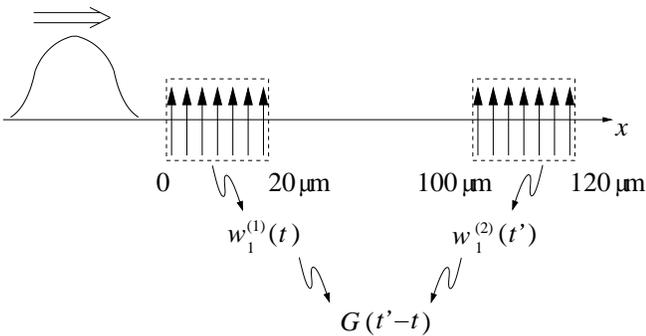, width=\columnwidth}
\caption{The first detector detects the particle at entry into the
  region of interest, the second one at exit. From the correlation of
  these two arrival-time distributions, the passage-time distribution
  is obtained. The $x$-positions correspond to the numerical examples
  discussed in Section \ref{numerical}.}
\label{passage_setup}
\end{center}
\end{figure}

In the present paper we investigate a particular measurement scheme
for passage times, mimicking the way a time-of-flight experiment is
typically performed: Employing a particular spin-boson detector model
for measuring quantum arrival times investigated in Ref.~\cite{hns1},
our measurement scheme involves two measurements of arrival time by
means of this detector, one upon entering the region of interest and
one when exiting; see Fig.~\ref{passage_setup}.
This scheme can be expected to distort the particle's wave function
less than a scheme based on the \hbox{(semi-)}con\-tin\-u\-ous
coupling of the particle to a measurement device. Indeed, it will be
seen that a passage-time measurement by means of ``slow'' detectors
yields a rather broad passage-time distribution. But detectors
responding quickly to the presence of the particle also yield a broad
distribution. This will be shown to be a quantum effect involving the
Heisenberg uncertainty relation.  However, there is an intermediate
range for the detector parameters yielding optimal results. In the
best case the precision of the measurement can be estimated to behave
like $E^{-3/4}$, improving the results obtained from clock models
\cite{ap1980, amm2003}.

In Sec.\ \ref{model} we review the detector model and its application
to quantum arrival-times. In Sec.\ \ref{reset-section} the particle's
wave function immediately after the detection is investigated, and in
Sec.\ \ref{passage-section} analytical formulas for the application of
the detector model to passage times are derived. Numerical examples
are investigated in Sec.\ \ref{numerical}. For simplicity, we
concentrate there on passage times for free particles; the extension
to passage times in the presence of barriers, and thus to tunneling
times, should proceed on similar lines. The precision of the
measurement scheme is estimated in Sec.~\ref{estimate}.

\section{The detector model and its arrival-time distribution}
\label{model}

We briefly review the detector model introduced in Refs. \cite{gs1990,
  lss1991, lss1997} and the arrival-time distribution obtained from
this model in Ref.\ \cite{hns1}. The detector consists of a
three-dimensional array of spins with ferromagnetic interaction in a
metastable state. The spins are weakly coupled to the environment,
modeled as a bath of bosons. The effect of the particle to be observed
is to strongly enhance the spin-bath coupling when the particle's wave
function overlaps with that of a spin. Thus when the particle is close
to a spin, this spin flips with strongly increased probability due to
the enhanced spin-bath coupling. By means of the ferromagnetic
interaction, this in turn triggers the subsequent spontaneous flipping
of the neighboring spins, and finally, by a domino effect, of all
spins, even in the absence of the particle. In this way, the single
spin flip induced by the presence of the particle is amplified to a
macroscopic event, and either the change in the detector state or in
the bath state can be measured.

The Hamiltonian for this model is
\begin{equation}
H = H_\mathrm{part} + H_\mathrm{det} +
H_\mathrm{bath} + H_\mathrm{spon} + H_\mathrm{coup} \,,
\label{hamiltonian}
\end{equation}
where we use the following definitions:
\begin{equation}
H_\mathrm{part} =
\frac{\hat{\mathbf{p}}^2}{2m}\,
\end{equation}
is the free Hamiltonian of the particle;
\begin{equation}
H_\mathrm{det} = \sum_j \frac{\hbar \omega_0^{(j)}}{2}
\hat{\sigma}_z^{(j)} - \sum_{j<k} \frac{\hbar \omega_J^{(jk)}}{2}
\hat{\sigma}_z^{(j)} \otimes \hat{\sigma}_z^{(k)}
\label{hdet}
\end{equation}
with $\excited_j$ denoting the excited state of the $j^\mathrm{th}$
spin and
\begin{equation}
\hat{\sigma}_z^{(j)} \equiv
\excited_{j\;j}\!\excitedbra \, - \, \ground_{j\;j} \! \groundbra \,,
\end{equation}
is the free Hamiltonian of the detector; $\hbar \omega_0^{(j)}$ is the
energy difference between ground state and excited state of the
$j^\mathrm{th}$ spin, and $\hbar \omega_J^{(jk)} \geq 0$ is the
coupling energy between the spins $j$ and $k$. Further,
\begin{equation}
H_\mathrm{bath} = \sum_{\pmb{\ell}} \hbar
  \omega_{\pmb{\ell}} \hat{a}_{\pmb{\ell}}^\dagger \hat{a}_{\pmb{\ell}},
\end{equation}
where $\hat{a}_{\pmb{\ell}}$ is the annihilation operator for a boson
with wave vector $\pmb{\ell}$, is the free Hamiltonian of the
environment, modeled as a bath of bosons, and
\begin{equation}
H_\mathrm{spon} = \sum_{j,\pmb{\ell}}\hbar \left(
    \gamma_{\pmb{\ell}}^{(j)} e^{i f_{\pmb{\ell}}^{(j)}}
    \hat{a}_{\pmb{\ell}}^\dagger \hat{\sigma}_-^{(j)} + \mathrm{h.c.}
  \right)
\label{spont}
\end{equation}
with
\begin{equation}
\hat{\sigma}_-^{(j)} \equiv \ground_{j\;j} \!
\excitedbra, \; \hat{\sigma}_+^{(j)} \equiv \left(
\hat{\sigma}_-^{(j)} \right)^\dagger = \excited_{j\;j} \! \groundbra,
\end{equation}
and the coupling constants $\gamma_{\pmb{\ell}}^{(j)}$ and the phases
$f_{\pmb{\ell}}^{(j)}$ depending on the particular realization of the
detector and the bath, describes the permanent spin-bath coupling.
The spin-bath coupling is strongly enhanced in the particle's
presence due to
\begin{equation}
H_\mathrm{coup}^{(j)} = \sum_j \chi^{(j)} \left(
    \hat{\mathbf{x}} \right) \sum_{\pmb{\ell}}\hbar \left(
    g_{\pmb{\ell}}^{(j)} e^{i f_{\pmb{\ell}}^{(j)}}
    \hat{a}_{\pmb{\ell}}^\dagger \hat{\sigma}_-^{(j)} + \hc \right),
\label{general_coupling}
\end{equation}
with
\begin{equation}
\left| g_{\pmb{\ell}}^{(j)} \right|^2 \gg \left|
  \gamma_{\pmb{\ell}}^{(j)}\right|^2;
\label{gamma_small}
\end{equation}
thus, the enhanced coupling of the $j^\mathrm{th}$ spin to the bath is
proportional to a sensitivity function $\chi^{(j)} (\mathbf{x})$ which
vanishes outside the region ${\mathcal G}^{(j)}$ where the
$j^\mathrm{th}$ spin is located. An example would be the
characteristic function which is $1$ on ${\mathcal G}^{(j)}$ and zero
outside. It will be assumed in the sequel that the Markov property
[see Eq.\ (\ref{markov})] holds in an appropriate continuum limit.
Initially, the system is prepared in the state
\begin{equation}
\ket{\Psi_0} = \nodetect \excitedall \ket{\psi_0},
\label{initial}
\end{equation}
where $\nodetect$ is the ground state of the bath (no bosons present),
and $\ket{\psi_0}$ denotes the spatial wave function of the particle.
For the $\hbar \omega_0^{(j)}$ only slightly above the energetic
threshold set by the ferromagnetic spin-spin coupling, and
$\gamma_{\pmb{\ell}}^{(j)}$ sufficiently small, the probability of a
spontaneous spin flip (false positive) is very small \cite{gs1990,
  lss1991, lss1997}. But when the particle is close to the
$j^\mathrm{th}$ spin, the excited state $\excited_j$ decays much more
quickly, due to the enhanced coupling, ``$g_{\pmb{\ell}}^{(j)}$,'' of the
spin to the bath.  Then, the ferromagnetic force experienced by its
neighbors is strongly reduced, and these spins can therefore flip rather
quickly, even in the absence of the particle, by means of the
$\gamma_{\pmb{\ell}}^{(j)}$. The first spin flip will be amplified to a
macroscopic event by the previously mentioned domino effect
\cite{gs1990, lss1991, lss1997}.

In Ref.\ \cite{hns1} we investigated, by means of the quantum
jump approach \cite{qu_jump}, the application of this detector model to
arrival-time measurements. The bath modes were indexed by wave vectors
\begin{equation}
\pmb{\ell} = \frac{2 \pi}{L_\mathrm{bath}} \left(
\begin{array}{c} n_1 \\ n_2 \\ n_3 \end{array} \right), \;\; n_i \in
\mathbb{N}_0,
\end{equation}
and the coupling constants were assumed to be of the form
\begin{eqnarray}
g_{\pmb{\ell}}^{(j)} & = & \left( \Gamma^{(j)}
    (\omega_\ell, \mathbf{e}_{\pmb{\ell}}) + {\mathcal O}
    (L_\mathrm{bath}^{-1}) \right)
\sqrt{\frac{\omega_{\ell}}{L_\mathrm{bath}^3}}, \nonumber \\
\gamma_{\pmb{\ell}}^{(j)} & = & \left( \Gamma_\mathrm{spon}^{(j)}
  (\omega_\ell, \mathbf{e}_{\pmb{\ell}}) +  {\mathcal O}
    (L_\mathrm{bath}^{-1}) \right)
\sqrt{\frac{\omega_{\ell}}{L_\mathrm{bath}^3}}
\label{coupling_constants_three}
\end{eqnarray}
with $\omega_\ell = c(\omega_\ell) \ell$, $\mathbf{e}_{\pmb{\ell}} =
\pmb{\ell}/\ell$, and
\begin{equation}
\label{gg}
\left| \Gamma^{(j)} (\omega_\ell, \mathbf{e}_{\pmb{\ell}}) \right|^2
\gg \left| \Gamma_\mathrm{spon}^{(j)} (\omega_\ell,
  \mathbf{e}_{\pmb{\ell}}) \right|^2.
\end{equation}
We introduce ``modified resonance frequencies''
\begin{equation}
\tilde{\omega}_0^{(j)} \equiv \omega_0^{(j)} -
  \sum_{{k=1}\atop{k\neq j}}^D
    \omega_J^{(jk)} ,
\label{modified}
\end{equation}
which incorporate the ferromagnetic spin-spin coupling. The
correlation function $\kappa_{\overline{g}g}^{(j)} (\tau)$ is defined
by
\begin{equation}
\kappa_{\overline{g}g}^{(j)} (\tau) \equiv \sum_{\pmb{\ell}}
\left| g_{\pmb{\ell}}^{(j)} \right|^2 e^{-i (\omega_\ell -
  \widetilde{\omega}_0^{(j)}) \tau},
\end{equation}
and similarly for $\kappa_{\overline{g} \gamma}^{(j)}$,
$\kappa_{\overline{\gamma} g}^{(j)}$, and $\kappa_{\overline{\gamma}
  \gamma}^{(j)}$. It is assumed that in the continuum limit,
$L_\mathrm{bath} \rightarrow \infty$, the correlation functions satisfy
the Markov property, i.e.,
\begin{equation}
  \kappa_{\overline{g}g}^{(j)} (\tau) \approx 0 \quad \mathrm{if}
  \quad \tau > \tau_\mathrm{c},
\label{markov}
\end{equation}
for some small correlation time $\tau_\mathrm{c}$, and likewise for
the other
correlation functions.
This is the case, e.g., if $\Gamma (\omega_\ell,
\mathbf{e}_{\pmb{\ell}})$ is independent of $\omega_\ell$, as in
quantum optics.

Let $\ket{\psi_\mathrm{cond}^t}$ denote the time evolution of the
spatial wave function under the condition that no boson is detected at
least until time $t$ (``conditional time evolution''). Then
\begin{equation}\label{nodet}
P_0(t) \equiv \left\langle \psi_\mathrm{cond}^t \right| \left. \!
\psi_\mathrm{cond}^t \right\rangle
\end{equation}
is the probability that no detection occurred until $t$, which is
decreasing in time, and
\begin{equation}\label{density}
w_1(t) \equiv - \frac{dP_0}{dt}(t)
\end{equation}
is the probability density for the first detection to occur at time
$t$. In Ref. \cite{hns1} it was shown that  $\ket{\psi_\mathrm{cond}^t}$ obeys a
``conditional Schr\"odinger equation'' of the form
\begin{equation}
i \hbar \frac{\partial}{\partial t} \ket{\psi_\mathrm{cond}^t} =
\Hcond \ket{\psi_\mathrm{cond}^t}
\end{equation}
with a Hamiltonian containing a complex potential,
\begin{equation}
\Hcond = 
\frac{\hat{\mathbf{p}}^2}{2m} + \frac{\hbar}{2}
\left[\delta_\mathrm{shift}(
\hat{\mathbf{x}}) -i A (
\hat{\mathbf{x}} ) \right]\,,
\label{three-d-schroedinger}
\end{equation}
where $\delta_\mathrm{shift}(
\hat{\mathbf{x}})$ and $A (\mathbf{x})$ are defined as follows.
$A (\mathbf{x})$ is a position-dependent detector decay rate,
\begin{eqnarray}
 A\left(\mathbf{x} \right) &=& 2\, \mathrm{Re}\, \sum_j
 \int_0^\infty d\tau \,
\{\kappa_{\overline{g}g}^{(j)}(\tau) \chi^{(j)}({\mathbf{x}})^2 +
\kappa_{\overline{\gamma}\gamma}^{(j)}(\tau)\} \nonumber \\ 
&=& \sum_j \left( \tilde{\omega}_0^{(j)} \right)^3
\left[\frac{c \left( \tilde{\omega}_0^{(j)} \right) -
    \tilde{\omega}_0^{(j)} c' \left( \tilde{\omega}_0^{(j)} \right)}{c
    \left( \tilde{\omega}_0^{(j)} \right)^4} \right]\nonumber\\
&& 
\phantom{\sum_j} \quad \times
 \int \frac{d \Omega_\mathbf{e}}{(2 \pi)^2}
\left( 
\left| \Gamma^{(j)} \left(
\tilde{\omega}_0^{(j)}, \mathbf{e} \right) \right|^2
\chi^{(j)}({\mathbf{x}})^2 \right. \nonumber \\
&& \phantom{\sum_j \quad \times \int \frac{d \Omega_\mathbf{e}}{(2
    \pi)^2}} \qquad \left.
 + \left| \Gamma^{(j)}_\mathrm{spon} \left(
\tilde{\omega}_0^{(j)}, \mathbf{e} \right) \right|^2 \right) \,,
\label{Athree} \nonumber \\
\end{eqnarray}
where the $d\Omega_\mathbf{e}$ integral is taken over the unit
sphere and where the contributions from
$\kappa^{(j)}_{\overline{g}\gamma}$ and
$\kappa^{(j)}_{\overline{\gamma}g}$ have been neglected, due to
(\ref{gg}). The terms have the familiar form of the Einstein
coefficients in quantum optics, where there would also be a sum over
polarizations. The real part of the potential, $\delta_\mathrm{shift}
\left(\mathbf{x}\right)$, is given by
\begin{equation}\label{3d-shift}
\delta_\mathrm{shift} \left(\mathbf{x}\right) = 
  2\, \mathrm{Im}\, \sum_j \int_0^\infty d\tau
  \{\kappa_{\overline{g}g}^{(j)}(\tau) \chi^{(j)}({\mathbf{x}})^2 +
  \kappa_{\overline{\gamma}\gamma}^{(j)}(\tau)\};
\end{equation}
in quantum optics, this would correspond to a line-shift.
Since the $\kappa_{\overline{\gamma}\gamma}$ term leads to a constant
it just gives an overall phase factor and can be omitted.

Note that the Hamiltonian on the r.h.s.\ of Eq.\
(\ref{three-d-schroedinger}) is not norm conserving due to the
imaginary contribution $-i \hbar A(\mathbf{x}) /2$, in accordance with
Eq. (\ref{nodet}).
 From Eqs.\ (\ref{three-d-schroedinger}) and (\ref{density}) one easily finds
\begin{eqnarray}
w_1 (t) & = & \frac{i}{\hbar} \left\langle \psi_\mathrm{cond}^t
  \left|\Hcond - \Hcond^\dagger \right| \psi_\mathrm{cond}^t
\right\rangle \nonumber\\
& = & \int d^3 x \; A \left( \mathbf{x} \right) \left|
  \left\langle \mathbf{x} \left| \psi_\mathrm{cond}^t
    \right. \right\rangle \right|^2,
\label{w}
\end{eqnarray}
which is an average of the position-dependent decay rate
$A(\mathbf{x})$ of the detector, weighted with the probability density
for the particle to be at position $\mathbf{x}$ and as yet undetected.

It was shown in Ref.\ \cite{hns1}, using an example with a
one-dimensional, simplified detector model, that $w_1(t)$ essentially
agrees with the probability density obtained from the discrete model
with 40 bath modes by means of standard unitary quantum mechanics, up
to the time of recurrences due to the discrete nature of the bath.

\section{The  reset operation after the first boson detection}
\label{reset-section}

\subsection{The particle reset state }

For the intended application of the detector model we will need more
than the time evolution up to the first detection. In particular, we
require knowledge of the ``reset state'' \cite{gch1993}, the particle
state immediately after the first boson detection.

Let the complete system (bath, detector, particle) be described at a
particular time by a density matrix of the form
\begin{equation}
\varrho \equiv \ket{0} \excitedall
\varrho_\mathrm{p}
\bra{\uparrow_1 \ldots \uparrow_D} \bra{0}\,,
\end{equation}
(with $\varrho_\mathrm{p}$ the particle density matrix), i.e., no
boson and all spins up. If a boson is found in a broadband boson
measurement a time $ \Delta t$ later, the density matrix for the
corresponding subensemble is obtained by sandwiching the above
expression with
\begin{equation}
\mathbb{P}_1 \equiv \sum_{\pmb{\ell}} \ket{1_{\pmb{\ell}}} \!
\bra{1_{\pmb{\ell}}},
\end{equation}
by the von Neumann-L\"uders projection rule~\cite{vN,Lueders}, and the
trace gives the probability. The subsystem consisting solely of the
particle is then described after the detection of a boson by a partial
trace,
\begin{equation}\label{partial}
\tr_\mathrm{det} \, \tr_\mathrm{bath} \; {\mathbb{P}_1}\, U(\Delta t, 0)  
  \varrho \; U^\dagger(\Delta t, 0)\,{\mathbb{P}_1}
\equiv {\mathcal R} \varrho_\mathrm{p} \; \Delta t \,,
\end{equation}
which defines the operation ${\mathcal R}$.
To calculate this we go to the interaction picture w.r.t.
\begin{eqnarray}
H_0 &\equiv& H - \left( H_\mathrm{coup} + H_\mathrm{spon} \right)
\nonumber\\
&=& H_\mathrm{part} + H_\mathrm{det} + H_\mathrm{bath}\nonumber\\
H_1&\equiv& H_\mathrm{coup} + H_\mathrm{spon}
\end{eqnarray}
and use standard second-order perturbation theory for $U_I(t,0) \equiv
\exp\{\idh H_0 t\}\,U(t,0)$. The zeroth order does not contribute to
Eq. (\ref{partial}), and neither does the first order since
${\mathbb{P}_1}$ acts once on $\ket{0}$. In second order only the term
with $H_1$ on the left and right survives, and in second order one
obtains after some calculation
\begin{widetext}
  \begin{eqnarray}
{\mathcal R} \varrho_\mathrm{p} \; \Delta t & = &
\tr_\mathrm{det} \, \tr_\mathrm{bath} \;
    {\mathbb{P}_1}\,e^{-\idh H_0\Delta t}
    \nonumber \\
& & \phantom{\tr_\mathrm{det} \,}
    \left( -\idh \right)
    \int\limits_0^{\Delta t}dt_1 e^{\idh H_0
      t_1}\;H_1e^{-\idh H_0 t_1} 
  \varrho \; \idh
    \int\limits_0^{\Delta t}dt_2 e^{\idh H_0
      t_2}H_1e^{-\idh H_0 t_2}e^{\idh H_0\Delta t}\,{\mathbb{P}_1}
    \nonumber \\
& = & 2\,\mathrm{Re} ~ e^{-\idh H_\mathrm{part}\Delta t}
\sum_{\pmb{\ell},j} \int\limits_0^{\Delta t} dt_1 
    \int\limits_0^{t_1}dt_2 \, e^{i 
      \left(\omega_{\ell} - \tilde{\omega}_0^{(j)}\right)( t_1 - t_2)}
\nonumber \\
& & \phantom{\sum \int dt_1 d_2} \quad \quad \times
    \left[\gamma_{\pmb{\ell}}^{(j)} + g_{\pmb{\ell}}^{(j)} \chi^{(j)}
      [\hat{\mathbf{x}}(t_1)] \right]  \, \varrho_\mathrm{p}
         \overline{\left[\gamma_{\pmb{\ell}}^{(j)} +
        g_{\pmb{\ell}}^{(j)} \chi^{(j)} [\hat{\mathbf{x}}(t_2)]
      \right]}e^{\idh H_\mathrm{part}\Delta t}, \nonumber \\
\label{reset_full}
\end{eqnarray}
\end{widetext}
where $\hat{\mathbf{x}} (t) = \hat{\mathbf{x}} + \hat{\mathbf{p}} t/m$
is the free time development of $\hat{\mathbf{x}}$ in the Heisenberg
picture. To obtain the second equality, the rectangular integration
area $t_1,~t_2 \in [0, \Delta t]$ has been split into two triangles,
$t_1 \in [0, \Delta t],~t_2\in [0,t_1]$ and $t_2 \in [0, \Delta
t],~t_1 \in [0, t_2]$.  The phases $f^{(j)}_{\pmb{\ell}}$ in $H_1$
have canceled since we are taking the trace over detector states.

\subsection{An example}
\label{example}

As an example we consider a simplified model with only one spatial
dimension and only one spin. This simplification is reasonable if the
radius of the region ${\mathcal G}_j$ is smaller than the distance
between spins. (Our assumption of locality of the interaction though
is a bit stronger than this however, since below, for calculational
convenience, we will extend the region ${\mathcal G}_j$ to a
half-line, i.e., $\chi(x) \to \Theta (x)$.) The vectors
$\mathbf{x},~\mathbf{p}$ and $\pmb{\ell}$ are replaced by $x,~p$ and
$\ell$, and there is of course no summation over the spin index $j$;
we will temporarily drop the superscript $^{(j)}$. The detector
Hamiltonian $H_\mathrm{det}$ [Eq.\ (\ref{hdet})] simplifies to
\begin{equation}
H_\mathrm{det}^1 = \frac{1}{2} \hbar \omega_0 \hat{\sigma}_z,
\end{equation}
and the modified resonance frequency $\widetilde{\omega}_0$ [Eq.\
(\ref{modified})] is replaced by the resonance frequency of the single
spin, $\omega_0$. Also, we will neglect $H_\mathrm{spon}$ in view of
the assumption $\left| \gamma_\ell \right|^2 \ll \left| g_\ell
\right|^2$. The time development in this model for a wave packet
incident from the left with detector and bath initially in state
$\ket{\uparrow~0}$ has been investigated, among other questions, in
Ref.\ \cite{hns1}.

We assume a maximal boson frequency $\omega_{_\mathrm{M}}$ and $N$ boson modes,
\begin{eqnarray}
\omega_\ell & = & \omega_{_\mathrm{M}} n/N, \quad n=1, \ldots, N
\nonumber \\
g_\ell & = & -i G \sqrt{\omega_\ell/N}.\label{ex}
\end{eqnarray}
We further take $\chi(x) = \Theta(x)$ where $\Theta$ denotes
Heaviside's step function.  As the particle we consider a cesium atom,
and the initial state at $t=0$ is assumed to be a pure state,
$\varrho_\mathrm{p} = \ket{\psi} \bra{\psi}$, with
\begin{equation}
\left\langle x \left| \psi \right.\right\rangle =
\sqrt{\frac{1}{\Delta x \sqrt{2 \pi}}} \; e^{-x^2/4(\Delta x)^2}
e^{i (mv_0/\hbar) x}.
\label{example_initial}
\end{equation}

With these simplifications the first line of Eq.\ (\ref{reset_full})
yields
\begin{widetext}
\begin{eqnarray}
\left\langle x \left| {\mathcal R} \varrho_\mathrm{p} \; \Delta t
  \right| x \right\rangle & = & \sum_\ell \left| g_\ell \right|^2
\int\limits_0^{\Delta t} dt_1 dt_2 \, e^{i (\omega_\ell - \omega_0)
  (t_1 - t_2)} \left\langle x \left| e^{-i H_\mathrm{part} \Delta t /
      \hbar} \Theta \left[ \hat{x}(t_1) \right] \right| \psi
\right\rangle \nonumber \\
& & \phantom{\sum_\ell \left| g_\ell \right|^2
\int\limits_0^{\Delta t} dt_1 dt_2 \, e^{i (\omega_\ell - \omega_0)
  (t_1 - t_2)}} \quad \quad \times
 \left\langle \psi \left|
    \Theta \left[ \hat{x} (t_2) \right] e^{i H_\mathrm{part} \Delta t
      / \hbar} \right| x \right\rangle \nonumber \\
& = & \sum_\ell \left| g_\ell \int\limits_0^{\Delta t} dt \, e^{i
    (\omega_\ell - \omega_0) t} \left\langle x \left| e^{-i
        H_\mathrm{part} (\Delta t - t) / \hbar} \Theta \left( \hat{x}
      \right) e^{-i H_\mathrm{part} t / \hbar} \right| \psi
  \right\rangle \right|^2.
\label{reset_example}
\end{eqnarray}
\end{widetext}
The contribution $\left\langle x \left| \cdots \right| \psi
\right\rangle$ has an intuitive explanation: The initial state evolves
freely until $t$ and is then projected onto the detector region. In an
intuitive picture, the time $t$ may be viewed as time of occurrence of
a boson. Since a boson can only be created when the spin couples to
the bath, the particle has to be inside the detector at $t$, hence the
projection. Then the state continues to evolve freely until $\Delta
t$. The integration is understood as sum over all possible ``paths''
satisfying the above picture, i.e., as sum over all times $t$.

\begin{figure}[ht]
\begin{center}
\epsfig{file=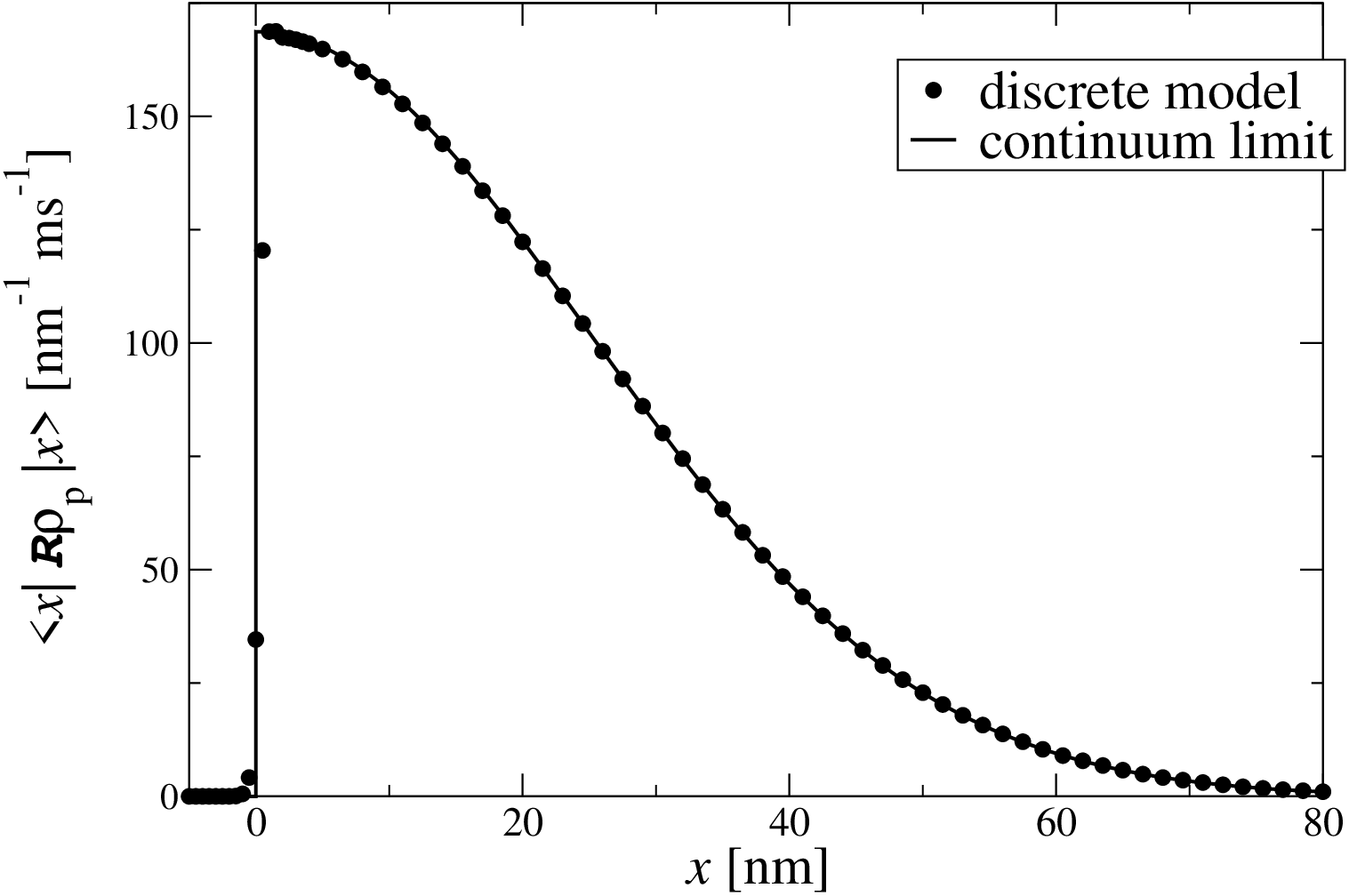, width=\columnwidth}
\caption{Dots: reset state $\left\langle x \left| {\mathcal R}
      \varrho_\mathrm{p} \right| x \right \rangle$ for a pure state
      $\varrho_\mathrm{p} = \ket{\psi} \bra{\psi}$ with $\ket{\psi}$ the
      Gaussian wave packet of Eq.~(\ref{example_initial}), $\Delta x = 50
      ~\mathrm{nm}$ and $v_0 = 1.79~\mathrm{m}/\mathrm{s}$; $\omega_0 =
 2.38
      \times 10^{12}~\mathrm{s}^{-1}$, $\omega_{_\mathrm{M}} = 4.6
 \,
      \omega_0$, $G = 2.782 \times 10^3~\mathrm{s}^{-1/2}$, $N=15$,
 and
      $\Delta t = 100 \; \omega_0^{-1} = 4.185 \times
      10^{-11}
~\mathrm{s}$. Solid line: reset state $A \, \left| \langle x
      \left| \Theta( \hat{x}) \right| \psi \rangle \right|^2$ from
      Eq.~(\ref{state}) for the corresponding continuum limit. Up to small
      deviations around $x=0$, the reset states from the discrete and the
      continuum model are in very good agreement.}
\label{reset_comp}
\end{center} \end{figure}

The second line of Eq.\ (\ref{reset_example}) can be evaluated, e.g.,
by inserting $\openone = \int_{-\infty}^\infty dk \ket{k} \bra{k}$
with momentum eigenfunctions $\langle x | k \rangle = \sqrt{1/2\pi}
\; e^{ikx}$ on the left of $\Theta (\hat{x})$, and
\begin{eqnarray}
\openone & = & \int_{-\infty}^\infty dx' \, dk' \, \ket{x'}
\left\langle x' \left| k' \right. \right\rangle \bra{k'} \nonumber \\
& = & \sqrt{\frac{1}{2\pi}} \int_{-\infty}^\infty dx'
\, dk' \,  e^{ik'x'} \ket{x'} \bra{k'}
\end{eqnarray}
on its right, and noting that
\begin{equation}
\left\langle k \left| \Theta \left( \hat{x} \right) \right| x'
\right\rangle = \sqrt{\frac{1}{2\pi}} \; e^{-ikx'} \; \Theta \left( x'
\right).
\end{equation}
A numerical illustration of $\left\langle x \left| {\mathcal R}
    \varrho_\mathrm{p} \right| x \right\rangle$ [note the division by
$\Delta t$ as compared to Eq.\ (\ref{reset_full}) or Eq.\
(\ref{reset_example})] for $N=15$ bosons modes is given in Fig.\
\ref{reset_comp} (dots).

\subsection{The continuum limit}

We return to the general expression in Eq.\ (\ref{reset_full}). For
simplicity we will assume in the following
\begin{equation} 
\chi^{(j)} ( \mathbf{x} ) \equiv \chi
( \mathbf{x}  ),~~~~~ j= 1, \cdots , D.
\label{j}
\end{equation}
In view of Eq.\ (\ref{gamma_small}) and the remarks after Eq.\
(\ref{initial}) we neglect the $\gamma_{\pmb{\ell}}^{(j)}$ terms. We
introduce the collective correlation function
\begin{equation} \kappa (\tau) \equiv \sum_j \sum_{\pmb{\ell}} \left|
    g^{(j)}_{\pmb{\ell}} \right|^2 e^{-i \left( \omega_\ell -
    \widetilde{\omega}_0^{(j)} \right) \tau}.\label{kappa}
\end{equation}
Since we assume that the coupling constants are such that the Markov
property holds in the continuum limit, i.e.,
\begin{equation}
\kappa (\tau) \approx
0~~~ \mathrm{if} ~~~\tau > \tau_c
\end{equation}
for some small correlation time
$\tau_c$, in the double integral of Eq.\ (\ref{reset_full}) only times
with $t_1 - t_2 \le \tau_c$ contribute, and if $\tau_c$ is small
enough one can write
\begin{equation} 
\chi [\hat{\mathbf{x}}(t_1)] \approx \chi [\hat{\mathbf{x}}(t_2)].
\label{approx}
\end{equation}
Then, with a change of variable, $\tau \equiv t_1-t_2$, the
integration over $\tau$ can be extended to $\infty$ if  $\tau_c \ll
\Delta t$. Specializing the definition Eq.\ (\ref{Athree}) to the case
at hand, we put 
\begin{eqnarray}
A \equiv 2\mskip 1mu \mathrm{Re} \int_0^\infty d\tau \,
\kappa (\tau)
\label{A}
\end{eqnarray}
and note that $A$ is given by the right side of Eq.\ (\ref{Athree})
without the terms $\Gamma^{(j)}_\mathrm{spon}$ and
$\chi^{(j)}({\mathbf{x}})$. One then obtains
\begin{widetext}
\begin{eqnarray}
{\mathcal R} \varrho_\mathrm{p} \; \Delta t &=& A\,
\int\limits_0^{\Delta t}dt_1 e^{-\idh H_\mathrm{part}(\Delta t-t_1)}
\chi (\hat{\mathbf{x}})e^{- \idh H_\mathrm{part} t_1}
  \varrho_\mathrm{p} e^{\idh H_\mathrm{part} t_1}
\chi(\hat{\mathbf{x}})e^{\idh  H_\mathrm{part}(\Delta t-t_1)} \nonumber \\
& = & A\, \chi(\hat{\mathbf{x}})
\varrho_\mathrm{p} \chi(\hat{\mathbf{x}})\; \Delta t +
\mathcal{O}(\Delta t^2).
\end{eqnarray}
\end{widetext}
 Hence, to first order in $\Delta t$, one  finally has 
\begin{eqnarray} \label{reset}
{\mathcal R} \varrho_\mathrm{p} = A\, \chi(\hat{\mathbf{x}})
\varrho_\mathrm{p} \chi(\hat{\mathbf{x}})
\end{eqnarray}
for the state  immediately after a detection. 

${\mathcal R}$ is called the reset operation. If $\varrho_\mathrm{p}$
is a pure state, then the reset state is also a pure state. In
particular, if $\varrho_\mathrm{p} =
\ket{\psi_\mathrm{cond}^t}\bra{\psi_\mathrm{cond}^t}$, then the reset
state is given by the wave function
\begin{equation} 
\ket{\psi_\mathrm{reset}^t} \equiv 
A^{1/2} \chi(\hat{\mathbf{x}}) \ket{\psi_\mathrm{cond}^t}.
\label{state}
\end{equation}
We note that this result is physically very reasonable since it means
that right after a detection of the particle by a detector located in
a specific region the particle is localized in that region. We also
note that [see Eq.\ (\ref{w})]
\begin{equation}
\left\langle \psi_\mathrm{reset}^t \left| \psi_\mathrm{reset}^t
  \right. \right\rangle = w_1(t),
\label{reset_norm}
\end{equation}
which is the probability density for the first detection.

We apply Eq.\ (\ref{state}) to the continuum limit of the example of
Subsection \ref{example}. In that one-dimensional and single-spin case
one has with Eqs.\ (\ref{ex}) and (\ref{kappa})
\begin{eqnarray}\label{excont}
  \kappa(\tau)&=& \omega_{_\mathrm{M}} |G|^2
  \sum_{n=1}^N\frac{1}{N}\frac{n}{N}e^{-i\omega_{_\mathrm{M}}\tau\,
    (n/N) + i\omega_0\tau}\\
  &\to& \omega_{_\mathrm{M}} |G|^2 \int_0^1 d\xi \,\xi \,
  e^{-i\omega_{_\mathrm{M}}\tau \, \xi+ i\omega_0\tau} \quad
  \mathrm{for}~ N~\to ~\infty.\nonumber
\end{eqnarray}
Thus one obtains in the continuum limit
\begin{eqnarray}
\kappa (\tau) & = & \frac{|G|^2}{\omega_{_\mathrm{M}}} \; \frac{
  \left( 1 + i \omega_{_\mathrm{M}} \tau \right) e^{-i
    (\omega_{_\mathrm{M}}-\omega_0) \tau} -
  e^{i \omega_0 \tau}}{\tau^2} \nonumber \\
A &=&  2\pi \left| G \right|^2
\,\frac{\omega_0}{\omega_{_\mathrm{M}}} \quad \mathrm{for}~
\omega_{_\mathrm{M}} > \omega_0 \nonumber \\
\delta_\mathrm{shift}&=& 2 \left|G\right|^2
\left(\frac{\omega_0}{\omega_{_\mathrm{M}}} \ln
  \left[\frac{\omega_0}{\omega_{_\mathrm{M}} - \omega_0} \right] -
  1\right),
\end{eqnarray}
and $\tau_\mathrm{c}$ is of the order of $\omega_0^{-1}$.  The
resulting spatial probability density $A \, \left| \langle x \left|
    \Theta ( \hat{x} ) \right| \psi \rangle \right|^2$ is plotted in
Fig.\ \ref{reset_comp} for the same initial state and same parameters
as in the discrete case. The plots are in very good agreement.

We note that in the present model with $f_{\pmb{\ell}}^{(j)}$
independent of $\mathbf{x}$ there is no explicit recoil on the
particle from the created boson. This is in line with the original
idea of a minimally invasive measurement. The absence of an explicit
recoil distinguishes the present detector model from other models
which are based on the direct interaction with the particle's internal
degrees of freedom. An example for such a model would be the detection
by means of laser induced fluorescence. In that case, the reset state
after the detection of the first fluorescence photon explicitely
incorporates a recoil due to the momentum of the emitted photon
\cite{gch2003}. It appears reasonable that in the present model no
such recoil on the particle occurs: After all, the boson is emitted
not by the particle but by the spin lattice. Hence, the recoil should
be experienced by this lattice, rather than by the particle, similar
to what occurs in the M\"ossbauer effect. Of course, the projection of
the wave packet onto the detector region by means of the reset
operation also changes the momentum distribution of the wave packet.

\subsection{Subsequent time development}

After detection of a boson, the further interaction of the particle
with the detector depends on the particular choice of parameters of
the detector model. The internal dynamics of the detector after the
first spin flip has been investigated in detail in Refs.\
\cite{lss1997, gs1990, lss1991}. Based on these results, several
choices are possible such that the amplification of the first spin
flip will not significantly change the spatial wave function after the
first spin flip, $\ket{\psi_\mathrm{reset}^t}$. This means that
effectively only one reset operation, associated with the very first
spin flip, has to be performed on the spatial wave function. Such
``minimally invasive'' detector models will be of interest if one is
interested in actual quantum mechanical limitations of a passage-time
measurement.

As a simple example, consider a ring of identical spins with
nearest-neighbor interaction,
\begin{equation}
\omega_0^{(j)} \equiv \omega_0 \quad \mathrm{and} \quad
\omega_J^{(jk)} \equiv \omega_J \; \delta_{j+1,k}
\end{equation}
with $\delta_{j+1,k}$ Kronecker's $\delta$, $\chi^{(j)} (\mathbf{x})
\equiv \chi (\mathbf{x})$ as before, and $j=D+1$ identified with
$j=1$. The rate for the neighbors of the first-flipped spin to flip
into \textit{their} ground state is denoted by $A_1 (\mathbf{x})$ and
is given by the right side of Eq.\ (\ref{Athree}), but with
$\widetilde{\omega}_0$ replaced by $\omega_0$ since the ferromagnetic
forces on these neighboring spins cancel. Choosing parameters as,
e.g., $\omega_0 \gg \widetilde{\omega}_0$ and
$\Gamma,~\Gamma_\mathrm{spon}$ independent of $\omega_\ell$ as well as
$c(\omega_\ell) \equiv c_0$, one has \hbox{$A_1 \gg A$}. By a kind of
domino effect the whole ring flips into the ground state; the mean
time needed for this given by $D/2A_1$ \cite{lss1997, gs1990,
  lss1991}. If this time is very short, as one can achieve by making
$\omega_0$ large (while $\widetilde{\omega}_0$ remains small to
prevent spontaneous spin flips before the first particle-induced spin
flip), the reset operations associated with these subsequent spin
flips will not significantly change the particle's state since the
wave function has been projected onto the detector by the first reset
operation already.

Another possibility to prevent the spatial wave function from being
changed by the amplification process would be to couple only one spin
to the particle, by choosing $g_{\pmb{\ell}}^{(j)} = g_{\pmb{\ell}}
\, \delta_{j,j_0}$ in Eq.\ (\ref{general_coupling}). (In fact, this
is the \emph{de facto} setup of the detector actually investigated in
Refs.\ \cite{gs1990, lss1991, lss1997}: Effectively only one spin
couples to the particle, and subsequently the other spins flip
spontaneously, i.e., without particle-enhanced spin-bath coupling.)

\section{Application to passage times}
\label{passage-section}
\subsection{General setup}
\label{general}

We now consider two detectors separated by some distance. As indicated
in the preceding section, we assume that the amplification of the
first spin flip to a macroscopic event is very fast and does not
change the spatial wave function. Thus, we take the probability
density $w_1(t)$ [Eq.\ (\ref{w})] for the first spin flip to be the
``measured arrival-time distribution'' and
$\ket{\psi_\mathrm{reset}^t}$ [Eq.\ (\ref{state})] as the particle's
state after the detection. Then, the joint probability density for the
first detector detecting the particle at $T$ \emph{and} the second one
detecting it at $T+\tau$ is given by
\begin{equation}
G(T, T+\tau) = w_1^{(1)} \left( T; \ket{\psi_0} \right) \;
w_1^{(2)} \left( \tau; \frac{\ket{\psi_\mathrm{reset}^T}}{\left\|
      \ket{\psi_\mathrm{reset}^T} \right\|} \right),
\end{equation}
where the superscripts indicate the detector under consideration and
the second argument is the initial state for which the respective
probability density is calculated. Since $w_1$ is bilinear in the wave
function [see Eq.\ (\ref{w})], this simplifies to
\begin{equation}
G(T, T+\tau) = w_1^{(2)} \left( \tau; \ket{\psi_\mathrm{reset}^T}
\right)
\end{equation}
by Eq.\ (\ref{reset_norm}).  The desired measured passage-time
distribution is then given by integration over the entry time $T$:
\begin{equation}
G(\tau) = \int dT \, w_1^{(2)} \left( \tau; \ket{\psi_\mathrm{reset}^T}
\right).
\end{equation}

\subsection{Numerical example}
\label{numerical}

\begin{figure}[ht]
\begin{center}
\epsfig{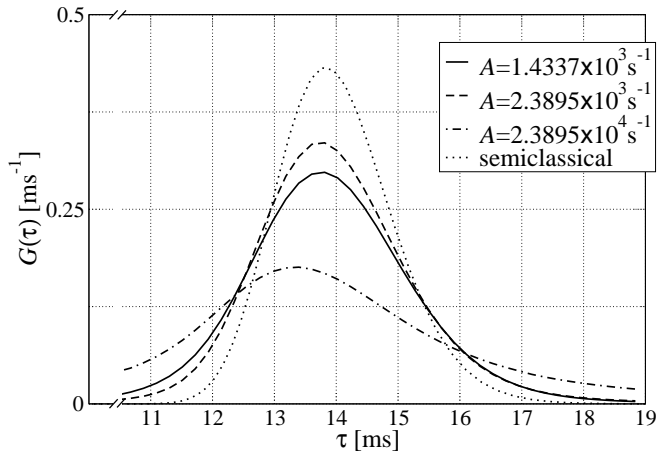}
\caption{Passage-time distributions calculated from the detector model
  in the continuum limit for three different values of the rate $A$
  for the first spin flip in the presence of the particle, all other
  parameters kept fixed; for comparison, the dotted line shows the
  passage-time distribution for an ensemble of classical particles
  which have the same momentum distribution as the initial wave
  packet.}
\label{passage_fig}
\end{center}
\end{figure}

In order to investigate basic features of a quantum passage-time
distribution obtained from the detector model in the continuum limit
and in the above measurement scheme, we consider as a simple example a
cesium atom in one dimension. The initial wave packet is prepared in
the remote past far away from the detector such that the free wave
packet (with no detectors present) would be described at $t=0$ by a
Gaussian minimal uncertainty packet around $x=0$ with $\Delta x = 1
\mu \mathrm{m}$ and average velocity $v_0 = 0.717 \mathrm{cm/s}$. Each
of the two identical detectors is described in the continuum limit by
an absorbing potential $-i V = -i \hbar A/2$ extending from $0$ to $20
\mu \mathrm{m}$ or $100 \mu \mathrm{m}$ to $120 \mu \mathrm{m}$,
resp., where we consider three different examples $A = 1.4337 \times
10^3~\mathrm{s}^{-1}, 2.3895 \times 10^3~\mathrm{s}^{-1}, 2.3895 \times
10^4~\mathrm{s}^{-1}$. These parameters are chosen in such a way that
transmission and reflection without detection, which are typical for
imaginary potentials \cite{ga1969} and have been extensively studied
in the framework of quantum arrival times \cite{dambo2002, hsm2003,
  bn2003b, dambo2003}, play no significant role.  Consequently, all
distributions shown in the following are normalized to 1 to a good
approximation. The passage-time distributions, calculated as described
in Section \ref{general}, are shown in Fig.~\ref{passage_fig}.

\begin{figure}[ht]
\begin{center}
\epsfig{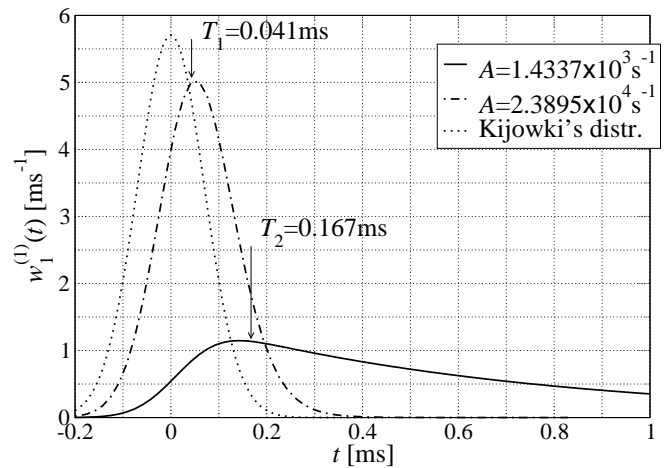}
\caption{Arrival-time distributions as obtained from the first
  detector. The detector with small $A$ gives a broad distribution
  indicating poor quality of this measurement. Enlarging $A$ carefully
  (to avoid reflection), one may approach Kijowski's axiomatically
  derived arrival-time distribution at $x=0$ (dotted line).}
\label{arrival_fig}
\end{center}
\end{figure}

It is seen that small and large values of $A$ give rise to rather
broad distributions, while the intermediate value yields a narrower
one. The reason for the broad distribution arising for small $A$ can
be understood by looking at the arrival-time distribution measured by
the first detector (see Fig.\ \ref{arrival_fig}): It is already this
distribution which is rather broad for small $A$.
Physically, small $A$ means that the detector is responding only
slowly to the presence of the particle; the undetected amplitude
$\ket{\psi_\mathrm{cond}^t}$ decays only slowly, yielding a broad
arrival-time distribution at each of the two detectors and
consequently a broad passage-time distribution. In other words, the
poor quality of the passage-time measurement arises from the poor
quality of the individual arrival-time measurements. The measurement
of the arrival time, however, can be improved by making $A$ large,
that is, by making the detector responding faster to the presence of
the particle. In Fig.\ \ref{arrival_fig} it is seen, e.g., that for $A
= 2.3895 \times 10^4~\mathrm{s}^{-1}$ one comes much closer to
Kijowski's arrival-time distribution; this in turn is known to have
minimum standard deviation among all those distributions fulfilling
certain axioms transferred to quantum mechanics from classical
arrival-time distributions \cite{jk1974}, and thus provides a sort
of ``ideal distribution.''

\begin{figure}[ht]
\begin{center}
\epsfig{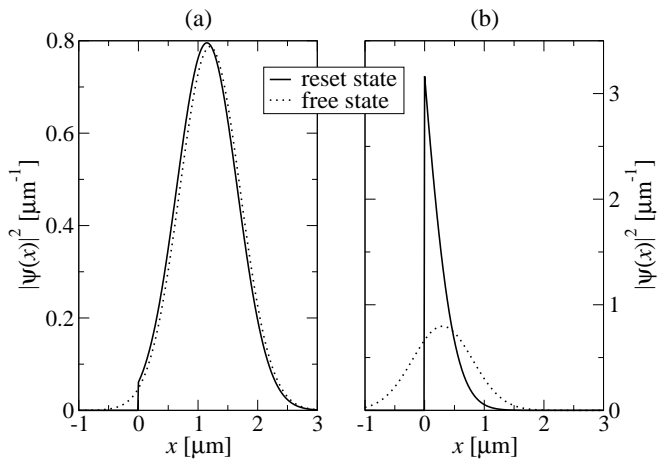}
\caption{The normalized reset state in position space immediately
  after a detection at (a) $T_2=0.167 \littlemathskip\mathrm{ms}$
  (with $A=1.4337 \times 10^3 \littlemathskip\mathrm{s}^{-1}$) and at
  (b) $T_1=0.041 \littlemathskip\mathrm{ms}$ (with $A=2.3895 \times
  10^4 \mathrm{s}^{-1}$), compared with the free wave packet at the
  respective instance of time (dotted line). The fast detection in
  case (b) has a strong impact on the wave function.}
\label{reset_x}
\end{center}
\end{figure}

\begin{figure}[ht]
\begin{center}
\epsfig{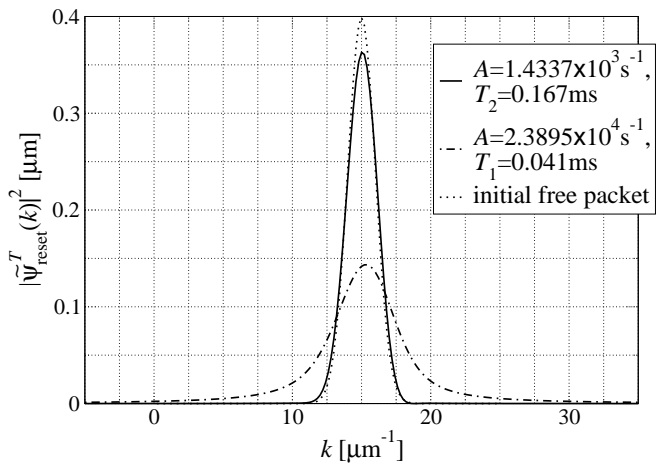}
\caption{Momentum distributions of reset states immediately after a
  detection at $T_2=0.167 \littlemathskip\mathrm{ms}$ (solid line,
  with $A=1.4337 \times 10^3 \littlemathskip\mathrm{s}^{-1}$) and at
  $T_1=0.041 \littlemathskip\mathrm{ms}$ (dash-dotted line, with
  $A=2.3895 \times 10^4 \littlemathskip\mathrm{s}^{-1}$), compared to
  the momentum distribution of the initial wave packet (dotted line).
  The fast detection in the latter case leads to a strong broadening
  of the momentum distribution.}
\label{reset_k}
\end{center}
\end{figure}

While the broad passage-time distribution for small $A$ can be
understood simply as due to the poor quality of the individual
measurements, the broad distribution for large $A$ is more
interesting. It can be understood looking at the reset state
immediately after the detection by the first detector. Of course, this
reset state depends on the instant of time when the detection took
place; as an example, we consider detection times $T_1 = 0.041
\littlemathskip\mathrm{ms}$ if $A=2.3895 \times 10^4\littlemathskip
\mathrm{s}^{-1}$ and $T_2 = 0.167 \littlemathskip\mathrm{ms}$ if
$A=1.4337 \times 10^3 \littlemathskip\mathrm{s}^{-1}$, which are close
to the maximum of the respective probability distribution $w_1(t)$
(see Fig.\ \ref{arrival_fig}). The reset states and the free packet at
the respective times are shown in Fig.\ \ref{reset_x}.
It is seen that the fast detection has a strong impact on the wave
function. Large $A$ means that $\ket{\psi_\mathrm{cond}^t}$, in
particular that part of $\ket{\psi_\mathrm{cond}^t}$ which overlaps
with the detector, decays very fast. But since the reset state
(\ref{state}) immediately after the detection is essentially the
projection of $\ket{\psi_\mathrm{cond}^t}$ onto the detector region,
the fast decay of this overlap yields a reset state which is very
narrow in position space, located at the very beginning of the
detector. Thus, by the Heisenberg uncertainty relation, it is very
broad in momentum space as can be seen in Fig.\ \ref{reset_k}. It is
intuitively clear that such a broad momentum distribution immediately
after the measurement of the entry time by the first detector yields a
broad passage-time distribution.  So the broad passage-time
distribution in case of large $A$ is due to the strong distortion of
the wave function by the first measurement. Since the broad momentum
distribution of the typical reset states in this case is enforced by
the Heisenberg uncertainty relation, this is a pure quantum effect.

\section{Width of the passage-time density}
\label{estimate}

In Refs.\ \cite{ap1980, amm2003} a measurement scheme for passage
times was investigated, based on the (semi-) continuous coupling of a
particle to a clock. It was argued in these references that the
precision of the measurement behaves like $E^{-1}$ where $E$ denotes
the kinetic energy of the particle. One now may wonder whether or not
the $E^{-1}$ behavior is a fundamental quantum limit for measuring
passage times. We argue that this is not the case since the present
measurement scheme by means of two spatially separated detectors
yields, for optimal parameter choices, passage-time densities with
widths behaving like $E^{-3/4}$. Thus, for low energies, we have an
example which breaks the $E^{-1}$ limitation of the clock model.

\subsection{Estimating the precision}

In this subsection we give an estimate for the width of the
passage-time density obtained from the present measurement scheme.  We
assume that the detectors can be described by the continuum limit and
that transmission and reflection without detection are negligible.
This assumption is justified in the examples of the preceding section,
which employed rectangular sensitivity functions.  It can also be
justified in general if one drops the restriction to rectangular
sensitivity functions $\chi(x)$ \cite{optpot}.

Considering particles with mean velocity $v_0$, the detector is
assumed to be constructed in such a way that the first detection
occurs with high probability in a spatial interval of length $L$.  The
length $L$ is related to $A$ of Eq.\ (\ref{A}), an average detection
rate, $L$ being of the order of $v_0/A$. The length $L$ imposes an
upper limit on the width of the reset state in position space,
\begin{equation}
\Delta x_\mathrm{reset}\leq L \approx v_0 /A.
\label{reset_x_bound}
\end{equation}
By the Heisenberg uncertainty relation, this immediately yields a
lower bound for the width $\Delta p_\mathrm{reset}$ of the reset
state in momentum space,
\begin{equation}
\Delta p_\mathrm{reset} \geq \hbar / 2\Delta x_\mathrm{reset} \geq
\hbar /2L.
\label{Dp}
\end{equation}
Note that this is only a very rough estimate, without taking into
account details of the actual wave packet. If the incident wave packet
is very narrow in position space, then $\Delta x_\mathrm{reset}$ also
may be much smaller than $L$, and consequently $\Delta
p_\mathrm{reset}$ may be much larger than $\hbar/2L$. Also, the reset
state may be far from being a Gaussian, and then already the first
inequality in Eq.\ (\ref{Dp}) may underestimate the width of $\Delta
p_\mathrm{reset}$ seriously.

Let $\Delta \tau$ denote the width of the measured passage-time
distribution. There are several contributions to this width. First, a
particle with velocity $v_0$ takes the time $\tau=d/v_0$ to travel the
distance $d$ between the two detectors, and therefore the width of the
reset state in position space contributes
\begin{equation}
\Delta \tau_{\mathrm{reset},x} = \Delta x_{\mathrm{reset}} / v_0
\label{Delta_tau_x}
\end{equation}
to $\Delta \tau$. Second, the width in momentum space contributes
according to
\begin{equation}
  \Delta \tau_{\mathrm{reset}, p} = \frac{ \Delta
    p_\mathrm{reset}}{mv_0} \frac{d}{ v_0} 
  \geq \frac{\hbar}{2\Delta x_\mathrm{reset}}\frac{d} {m v_0^2}.
\label{Delta_tau_p}
\end{equation}
Third, the width of the delay, $\Delta \tau_\mathrm{delay}$, for the
first detection due to the spin-boson interaction is roughly
\begin{equation}
\Delta \tau_\mathrm{delay} = 1/A \approx L/v_0.
\label{delay}
\end{equation}
An estimate for the width of the passage-time distribution is given by
the sum of these contributions, where $\Delta \tau_\mathrm{delay}$ has
to be counted twice since it arises in both detectors:
\begin{eqnarray}
  \Delta \tau &=& 2 \Delta \tau_\mathrm{delay} + \Delta
  \tau_{\mathrm{reset}, x} + \Delta \tau_{\mathrm{reset},p}\nonumber\\
  &\gtrsim& 2/A + \Delta
  x_\mathrm{reset}/v_0+ \hbar d/2mv_0^2\Delta x_\mathrm{reset} .
\label{Delta_tau}
\end{eqnarray}
From this estimate it is again seen that both small as well as large
values of $A$, i.e., both slow as well as fast detectors, lead to
rather broad passage-time distributions (due to $\Delta
\tau_\mathrm{delay} \sim 1/A$ and $\Delta \tau_{\mathrm{reset}, p}
\sim 1/ \Delta x_\mathrm{reset} \sim 1/L\sim A$, respectively).

\subsection{Optimal parameters}

Having established the general estimate for $\Delta \tau$ in Eq.\
(\ref{Delta_tau}), we now turn to the task of finding optimal
parameters, minimizing $\Delta \tau$. We are interested in measuring
the passage time through a spatial interval of length $d$, the
distance between the starting points of the two detectors, which we
regard as fixed. First, we consider given detectors, i.e., a given
detection rate, $A$. Again, particles with mean velocity $v_0$ would
be detected within an interval with length $L$ given approximately by
$L \approx v_0 / A$. This means that the velocities must not be too
large since in order to avoid undetected transmission $L$ must not
exceed the actual length of the detector (and in particular $L$ must
not exceed $d$). Quantum effects, however, are expected to play a role
for slow particles while fast particles may be treated classically,
hence this is not a serious drawback.

We assume that the reset state is a Gaussian wave packet, or at least
close to a Gaussian, so that in the second line of Eq.\
(\ref{Delta_tau}) the approximate equality holds,
\begin{equation}
\Delta \tau \approx 2/A + \Delta
x_\mathrm{reset}/v_0+ \hbar d/2mv_0^2\Delta x_\mathrm{reset}.
\label{Delta_tau_gauss}
\end{equation}
We note that $2/A$ is a purely detector-related quantity, while the
remaining contribution to $\Delta \tau$ is determined by the shape of
the reset state only. We will first optimize this latter quantity. For
given particles with given mean velocity, i.e., $m$ and $v_0$ fixed,
this is minimal for
\begin{equation}
\Delta x_\mathrm{reset}^\mathrm{opt} = \sqrt{\hbar d / 2mv_0}.
\label{Delta_x_opt}
\end{equation}
Substituting this into Eq.\ (\ref{Delta_tau_gauss}) and writing
$E=mv_0^2 / 2$ for the kinetic energy of the incident particles yields
\begin{equation}
\Delta \tau_\mathrm{opt;~reset} \approx 2/A + \sqrt{\hbar d \sqrt{m/2}}
\; E^{-3/4}.
\label{Delta_tau_opt_reset}
\end{equation}
The subscript ``$\mathrm{opt;~reset}$'' indicates that only the reset
state was optimized while the detection rate $A$ was considered as a
given quantity.

Aiming at optimizing also the detection rate $A$ for minimal $\Delta
\tau$, one would like to choose $A$ as large as possible in order to
reduce the $2/A$ contribution to $\Delta \tau$. However, one has to
take into account that the width of the reset state is bounded by Eq.\
(\ref{reset_x_bound}). Thus, given $d$, $m$, and $v_0$, the decay rate
must be at most of the order of $v_0/\Delta
x_\mathrm{reset}^{\mathrm{opt}}$ with $\Delta
x_\mathrm{reset}^\mathrm{opt}$ as in Eq.\ (\ref{Delta_x_opt}). In
fact, we may choose the incoming state such that it forms a Gaussian
minimal uncertainty packet at the starting point of the first detector
with width in position space
\begin{equation}
\Delta x = \Delta x_\mathrm{reset}^\mathrm{opt} = \sqrt{\hbar d / 2mv_0},
\end{equation}
and choose further
\begin{equation}
A \approx v_0 / 2 \Delta x = \sqrt{m / 2 \hbar d} \; v_0^{3/2}.
\label{A_opt}
\end{equation}
We note that, by this parameter choice, yet another requirement of a
good measurement scheme is fulfilled: The detection by the first
detector will not change the wave function too strongly. The reset
operation after this detection is essentially a projection onto the
detector region, and the detection is slow enough that at typical
detection times most of the wave packet overlaps with the detector,
hence the projection does not change the wave function too much. Thus,
the reset state will be close to a Gaussian wave packet with width
$\Delta x = \Delta x_\mathrm{reset}^\mathrm{opt}$. Substituting Eq.\
(\ref{A_opt}) into Eq.\ (\ref{Delta_tau_opt_reset}) finally yields
\begin{equation}
\Delta \tau_\mathrm{opt} \approx \sqrt{5 \hbar d \sqrt{m/2}}
\; E^{-3/4}.
\end{equation}

We stress that, independent of the detection rate $A$, the optimal
energy dependence of $\Delta \tau$ is limited to $E^{-3/4}$ already by
the dependence of $\Delta \tau$ on the width of the reset state in
position space, $\Delta \tau_{\mathrm{reset}, x} \sim \Delta
x_\mathrm{reset}$ [Eq.\ (\ref{Delta_tau_x})], and on its width in
momentum space, $\Delta \tau_{\mathrm{reset}, p} \sim \Delta
p_\mathrm{reset} \sim 1/\Delta x_\mathrm{reset}$ [see Eq.\
(\ref{Delta_tau_p})].

For the example of a cesium atom with $v_0 = 0.717 \littlemathskip
\mathrm{cm/s}$ and a distance of $d = 100
\littlemathskip\mu\mathrm{m}$ between the detectors (with rectangular
sensitivity function) of the preceding section, optimal values would
be according to Eqs.\ (\ref{Delta_x_opt}) and (\ref{A_opt})
\begin{equation}
  \Delta x_\mathrm{opt} = 1.83 \littlemathskip\mu \mathrm{m},
  \;\; \text{and} \;\;
  A_\mathrm{opt} = 1.959 \times 10^3 \littlemathskip\mathrm{s}^{-1}.
\end{equation}
Considering the wave packet with $\Delta x = 1 \littlemathskip\mu
\mathrm{m}$ actually investigated in the examples, the optimal decay
rate according to Eq.\ (\ref{A_opt}) would be $A_\mathrm{opt} (\Delta
x = 1 \littlemathskip\mu \mathrm{m}) = 3.585 \times 10^3
\littlemathskip\mu \mathrm{s}^{-1}$.  This is consistent with the
observation that the example with $A$ closest to $A_\mathrm{opt}
(\Delta x = 1 \littlemathskip\mu \mathrm{m})$ yields the narrowest
distribution.

\section*{Summary}

We have investigated the continuum limit of a fully quantum mechanical
spin-boson model for the detection of a moving particle and its
application to passage-time measurements. The continuum limit has been
derived under the condition that the spin-boson interaction satisfies
the Markov property, an assumption that was explicitly verified in
Ref.\ \cite{hns1}.  Analytical expressions for the state immediately
after the first detection have been obtained, and for an example with
a simplified detector model and 15 boson modes it was shown
numerically that the continuum limit is in very good agreement with
the discrete model. Further, analytical expressions for the
passage-time distribution have been obtained, and numerical examples
for passage-time measurements have been discussed.  Detectors with a
very slow response yield broad distributions, due to the poor quality
of the individual measurements, and so do very fast detectors, due to
the strong distortion of the particle's wave function by the
measurement.  Intermediate detectors yield narrower distributions. The
optimal precision of the present measurement scheme has been estimated
to behave like $E^{-3/4}$, where $E$ is the kinetic energy of the
incident particle. For slow particles this is better than, and in
contrast to, a scheme based on quantum clocks which yields an $E^{-1}$
behavior.

\section*{Acknowledgment}

This work was supported in part by NSF Grant PHY 0555313.


\end{document}